  \documentclass[aps,preprint,preprintnumbers, amsmath, amssymb, prb,showpacs]{revtex4-1}
\usepackage{times}
\usepackage{graphicx}
\usepackage{psfrag}
\usepackage{ae}
\usepackage{amsmath,amssymb}
\usepackage[usenames]{color}

\begin{document}
\date{\today}

 \author{Jos\'e Rafael Bordin} 
\email{josebordin@unipampa.edu.br}
\affiliation{Campus Ca\c capava do Sul, Universidade Federal
do Pampa, Caixa Postal 15051, CEP 96570-000, 
Ca\c capava do Sul, RS, Brazil}

\author{Leandro B. Krott} 
 \email{leandro.krott@ufrgs.br}
 \affiliation{Instituto 
 de F\'{\i}sica, Universidade Federal
 do Rio Grande do Sul\\ Caixa Postal 15051, CEP 91501-970, 
 Porto Alegre, RS, Brazil}
 
 \author{Marcia C. Barbosa} 
 \email{marcia.barbosa@ufrgs.br}
 \affiliation{Instituto 
 de F\'{\i}sica, Universidade Federal
 do Rio Grande do Sul\\ Caixa Postal 15051, CEP 91501-970, 
 Porto Alegre, RS, Brazil}

\title{High pressure induced phase transition and superdiffusion in anomalous fluid confined in flexible 
nanopores}	

\begin{abstract}

The behavior of a confined spherical symmetric anomalous fluid under high external pressure was
studied with Molecular Dynamics simulations. The fluid is modeled by a core-softened
potential with two characteristic length scales, which in bulk reproduces the dynamical, thermodynamical
and structural anomalous behavior observed for water and other anomalous fluids.
Our findings show that this system has a superdiffusion regime
for sufficient high pressure and low temperature. As well, our results 
indicate that this superdiffusive regime is strongly related with
the fluid structural properties and the superdiffusion to diffusion transition is a first order phase transition. 
We show how the simulation time and statistics are important to obtain the correct dynamical behavior of the 
confined fluid. Our results are discussed in the basis of the two length scales.

\end{abstract}


\maketitle
\section{Introduction}

The dynamical  behavior of fluids at nanoscale has been attracting attention 
recently. The reason behind this attention
is that the liquids under nano confinement
exhibit  hydrodynamic, dynamic and structural
properties different from the mesoscopic
confined and bulk systems~\cite{Tabeling14}.
For instance, the fast flow of the liquids confined in
nano structures
 can not be described 
by the classical hydrodynamic~\cite{Majumder05,Holt06,Wh08}.
This inconsistence between the experiments and 
the classical theories become even more 
significative in anomalous liquids, such as water~\cite{Qin11,Lee12}, where
the enhancement flow is higher than the enhancement observed
in other fluids~ \cite{Majumder05,Holt06,Wh08}.
In the particular case of water, this unusual dynamics
might lead  to important technological applications in 
desalinization~\cite{Tabeling14,Holt06,Wh08}.

For anomalous liquids, as water, the confinement
produces additional properties such as: the
presence of a well defined layered structure~\cite{Nanok09},   crystallization 
of the contact layers
at high temperatures~\cite{Cu01, JCP06, Alabarse12}, the increase
of the diffusion coefficient with the increase
of the confinement~\cite{Farimani11,St06}  and the oscilatory behavior in
the 
superflow~ \cite{Jakobtorweihen05,Chen06b,Qin11}.

What are the bulk  properties in anomalous fluids that
under confinement might lead to the appearance of unusual dynamics effects?
Most liquids contract upon cooling. This is not the case of the
anomalous liquids. For
them the specific volume at ambient pressure starts to increase 
when cooled below a certain temperature. In addition, while
most  liquids diffuse faster as pressure and density decreases
and contract on cooling, anomalous liquids exhibits a maximum density
 at constant pressure, and the
diffusion coefficient increases under compression~\cite{URL}.
The most well know anomalous fluid is water~\cite{Ke75,An76,Pr87},
but Te~\cite{Th76}, Ga,
Bi~\cite{Handbook}, Si~\cite{Sa67,Ke83}, $Ge_{15}Te_{85}$~\cite{Ts91},  liquid 
metals~\cite{Cu81}, graphite~\cite{To97}, silica~\cite{An00,Sh02,Sh06}, silicon~\cite{Sa03},  
$BeF_2$~\cite{An00} also show the presence of 
thermodynamic anomalies~\cite{Pr87}. In addition to water~\cite{Ne01, Ne02a}, 
silica~\cite{Sa03,Sh02,Sh06,Ch06} and silicon~\cite{Mo05} 
exhibit a maximum in the 
diffusion coefficient at constant temperature. 
As well, colloidal systems and globular proteins can also exhibit
anomalous properties~\cite{Vilaseca11}.

The diffusion coefficient, in bulk, is obtained from the scaling
factor between the mean square displacement and the 
exponent of the time, namely $<r(0)r(t)> = 2D t^{\alpha}$. For the anomalous liquids
in the bulk this scale factor follows the Fick 
diffusion. That means that the mean square displacement
is linear with time, $\alpha=1$. As the system becomes 
confined in addition to 
the fickian dynamics two anomalous no-fickian behaviors are observed.
The first, the superdiffusive regime, complies all
the $\alpha>1$ cases with a 
 fast dynamics. The limit is an ideal
system where the molecules can move with constant
velocity and, therefore, ballistic diffusion, $\alpha=2$.
The second, the subdiffusive regime, includes dense systems in which
the dynamics is slower and the particles move  in a chain-like structure
and cannot pass each other forming a single-file diffusion with
$\alpha=0.5$. The transition between these regimes
was observed in fluids confined inside nanotubes, and
depends on the radius and length of the nanotube as well
as on the time of observation of the moviment~\cite{Zheng12,Farimani11}. 
It is not clear, however, how the dynamics is related
with the structure.

In this paper we explore the connection
between the dynamic and structural 
anomalous behavior in nanoconfined systems
suggesting that the layering structure governs the 
dynamic transition. We propose that for
very high pressures the transition between
fickian to superdiffusive is related with
the structural transition between two to three layers.

The fluid is modeled using a two length
scale potential. 
Coarse graining potentials are a suitable tool to investigate 
the properties of a general confined anomalous fluids. Recently 
we have shown that this effective potential
is capable to reproduce the enhancent flow
and the high diffusion coefficient of nanoconfined
anomalous fluids~\cite{Zheng12,Farimani11}. For small pressures
the structure is related to thermodynamic phase
transitions in the wall~\cite{Bordin14a,Krott13b}.
Our model in the bulk
exhibits the thermodynamic, dynamic and structural anomalous behavior
observed in anomalous fluids in bulk~\cite{Oliveira06a,Oliveira06b} 
and in confinement~\cite{Krott13a,Krott13b,Krott14a,Bordin12b,Bordin13a, Bordin14b}.  The paper is 
organized as follows: in Sec. II we introduce the model
and describe the methods and simulation details; the results are given in Sec. III; and 
in Sec. IV we present our conclusions.

\section{The Model and the Simulation details}
\label{Model}

\subsection{The Model}

 \begin{figure}[ht]
 \begin{center}
 \includegraphics[width=8cm]{fig1a.eps}
 \includegraphics[width=8cm]{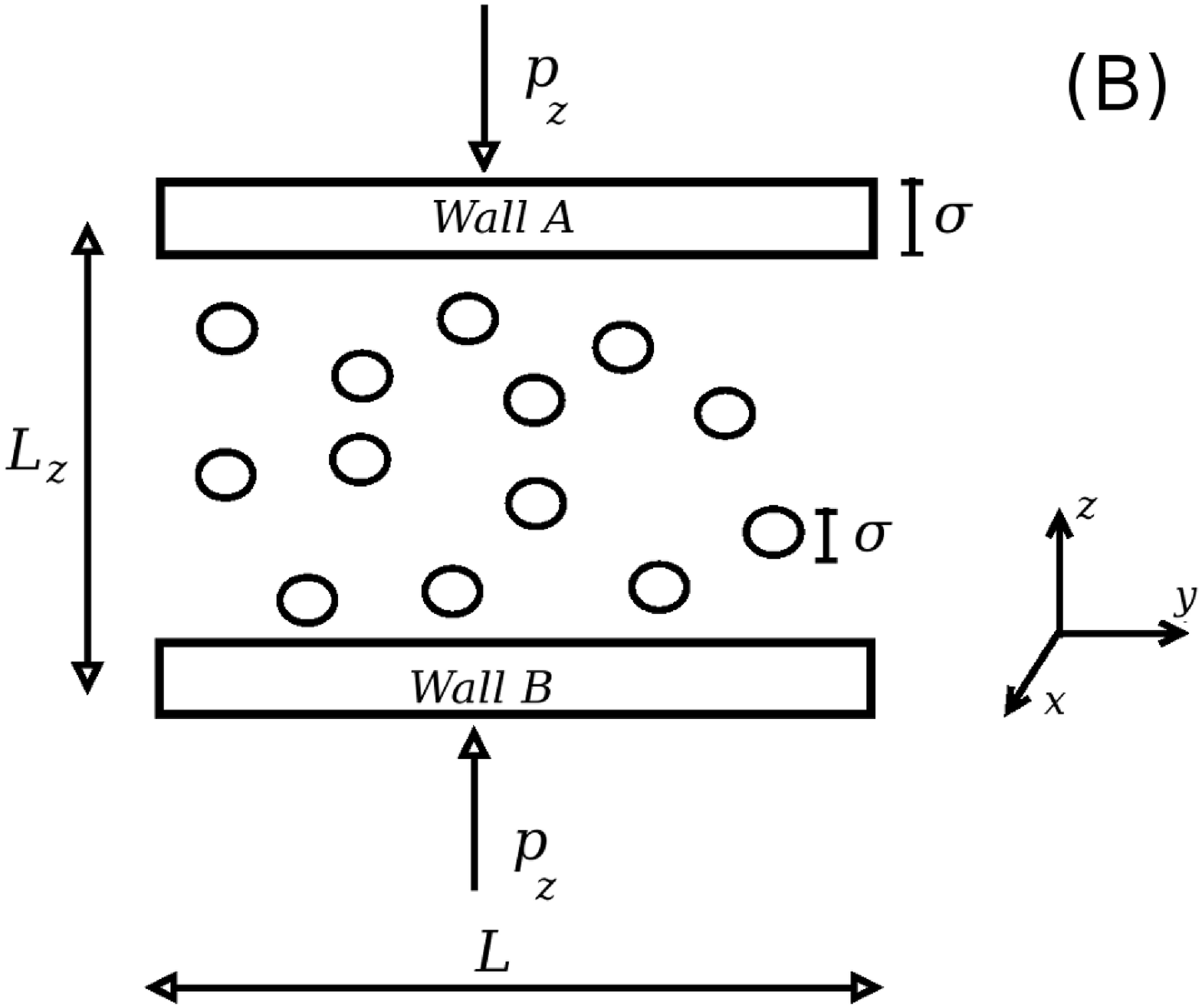}
 \end{center}
 \caption{(A) Interaction potential as function of the particles  separation. 
 (B) Schematic depiction of the simulation cell with the fluid and walls. The walls
 are separated by a distance $L_z$, have thickness $\sigma$ and an external pressure $p_z$ is applied in the
 $z$ direction.}
 \label{fig1}
 \end{figure}

 The anomalous fluid was modeled as spherical core-softened particles with mass $m$ and 
 effective diameter $\sigma$. The interaction is obtained by the potential~\cite{Oliveira06a}
 \begin{equation}
 \frac{U(r_{ij})}{\varepsilon} = 4\left[ \left(\frac{\sigma}{r_{ij}}\right)^{12} -
 \left(\frac{\sigma}{r_{ij}}\right)^6 \right] + u_0 {\rm{exp}}\left[-\frac{1}{c_0^2}\left(\frac{r_{ij}-r_0}{\sigma}\right)^2\right]
 \label{AlanEq}
 \end{equation}
 where $r_{ij} = |\vec r_i - \vec r_j|$ is the distance between the two fluid particles $i$ and $j$.
 This equation has two terms: the first one is the standard 12-6 Lennard-Jones (LJ)
 potential~\cite{AllenTild} and the second one is a gaussian
 centered at $r_0$, with depth $u_0$ and width $c_0$.
 Using the parameters $u_0 = 5.0$, $c = 1.0$ and $r_0/\sigma = 0.7$ this equation 
 represents a two length scale potential, with one scale 
 at  $r_{ij}\equiv r_1\approx 1.2 \sigma$, where the 
 force has a local minimum, and the other scale at  $r_{ij}\equiv r_2 \approx 2 \sigma$, where
 the fraction of imaginary modes has a local minimum~\cite{Oliveira10}.
 The potential is shown in Fig.~\ref{fig1}(A).
 Despite the mathematical simplicity this model 
 exhibits the bulk water-like anomalies~\cite{Oliveira06a, Oliveira06b, Kell67,Angell76}
 as well confined water properties~\cite{Bordin12b, Bordin13a, Krott13a, Krott13b, Krott14a, Bordin14a, Bordin14b}.

 Was already shown that the confined fluid properties are strongly affected by the 
 nanopore mobility~\cite{Krott13b, Bordin14a, Bernardi10,Yong13}. Since we want to fix the pressure at high values,
 we explore behavior of this anomalous fluid confined in a non-rigid nanopore~\cite{Krott13b, Bordin14a, Bernardi10,Yong13}.
The nanopore was modeled as two parallel flat plates. The simulation box is a parallelepiped 
with dimensions $L_x\times L_y\times L_z$. 
 The model for the fluid-wall system is illustrated in Fig.~\ref{fig1}(B).
 Two walls, A in the top and B in the  bottom, are placed in the limits of the $z$-direction 
 of the simulation box. The sizes $L_x$ and $L_y$ are fixed in all simulations,
 and defined as $L_x = L_y = L = 40\sigma$. The values of $L_z$ depends on the
 applied pressure $p_z$ in the $z$-direction. The system was modeled in the $NLp_zT$ ensemble
 using the Lupkowski and van Smol method of fluctuating confining walls~\cite{LupSmol90}
 to fix $p_z$.
 
 The walls are flat and purely repulsive, and the interaction between a fluid 
 particle and these walls is represented by the Weeks-Chandler-Andersen (WCA)~\cite{WCA71} potential,
 \begin{equation}
 \label{LJCS}
 U^{\rm{WCA}}(z_{ij}) = \left\{ \begin{array}{ll}
 U_{{\rm {LJ}}}(z_{ij}) - U_{{\rm{LJ}}}(z_c)\;, \qquad z_{ij} \le r_c\;, \\
 0\;, \qquad \qquad \qquad \qquad \quad z_{ij}  > r_c\;.
 \end{array} \right.
 \end{equation}
 Here, $U_{{\rm {LJ}}}$ is the standard 12-6 LJ potential, included in the first term of Eq.~(\ref{AlanEq}),
 and $r_c = 2^{1/6}\sigma$ is the usual cutoff for the WCA potential. Also, the term $z_{ij}$ measures the distance
 between the wall at $j$ position and the $z$-coordinate of the fluid particle $i$.

\subsection{The simulation details}

The physical quantities are computed
 in the standard LJ units~\cite{AllenTild},
\begin{equation}
\label{red1}
r^*\equiv \frac{r}{\sigma}\;,\quad \rho^{*}\equiv \rho \sigma^{3}\;, \quad 
\mbox{and}\quad t^* \equiv t\left(\frac{\epsilon}{m\sigma^2}\right)^{1/2}\;,
\end{equation}
for distance, density of particles and time , respectively, and
\begin{equation}
\label{rad2}
p^*\equiv \frac{p \sigma^{3}}{\epsilon} \quad \mbox{and}\quad 
T^{*}\equiv \frac{k_{B}T}{\epsilon}
\end{equation}
for the pressure and temperature, respectively. 
Since all physical quantities are defined in reduced LJ units in this paper, 
the $^*$ is  omitted, in order to simplify the discussion.

The simulations were performed
at constant number of particles, constant $L$, constant perpendicular pressure and 
constant temperature ($NLp_{z}T$ ensemble). The perpendicular 
pressure was fixed
using the Lupkowski and van Smol method~\cite{LupSmol90}. 
In this technique, the nanopore walls had  
translational freedom in the $z$-direction,
acting like a piston in the fluid, and a constant force controls the 
pressure applied in the confined direction.
This scenario is similar to some recent experiments on
water confined inside nanopores at externally applied
high pressures~\cite{Alabarse14, Catafesta14}.
Considering the nanopore flexible walls,
the resulting force in a fluid particle is then
\begin{equation}
 \vec F_R = -\vec\nabla U + \vec F_{iwA}(\vec r_{iA}) + \vec F_{iwB}(\vec r_{iB})\;,
\end{equation}
where  $\vec F_{iwA(B)}$ indicates the interaction 
between the particle $i$ and the wall $A(B)$. 
Since the walls are non-rigid and time-dependent, we have to solve the equations of motion
for $A$ and $B$,
\begin{equation}
 m_w\vec a_{A(B)}= p_{z}S_w\vec n_{A(B)} - \sum_{i=1}^N \vec F_{iwA(B)}(\vec r_{iA(B)}),
\end{equation}
where $m_w$ is the piston mass, $p_{z}$ is the applied pressure in the 
system, $S_w$ is the wall area and $\vec n_A$ is an unitary vector 
in positive $z$-direction, while $\vec n_B$ is a negative unitary vector. Both 
pistons ($A$ and $B$) have mass $m_w=m$, width $\sigma$ and area equal to $S_w = L^2$.

 The system temperature was fixed using the Nose-Hoover heat-bath with a coupling parameter $Q = 2$
 and was varied from small temperatures, $T=0.01$ to higher temperatures $T=0.4$. 
  Standard periodic boundary conditions were applied in the $x$ and $y$ 
 directions. The equations of motion for the fluid particles and the walls
 were integrated using the velocity Verlet algorithm,
 with a time step $\delta t = 0.001$.
 
 Five independent runs were performed to evaluate the properties of the
 confined fluid. The initial system was generated placing the fluid
 particles randomly in the space between the walls. The initial displacement
 for the simulations was $L_{z0} = 15$.
 We performed $5\times10^5$ steps to equilibrate the system.
  This equilibration time
 was taken in order to ensure that the walls reached the equilibrium position for the fixed 
 values of $p_z$. These steps are then followed 
 by $1\times10^8$ steps for the results production stage. The large production time
 is necessary to observe the correct dynamical behavior of the confined fluid.
 
 The fluid-fluid interaction, Eq.~(\ref{AlanEq}), has a cutoff radius $r_{\rm cut}/\sigma = 3.5$.
 The number of particles was fixed in $N = 1000$, and four values of pressure where simulated:
 $p_z = 7.0$, 8.0, 9.0 and 10.0.
  Due to the excluded volume originated by the nanopore-fluid interaction, the distance $L_z$ 
between the walls needs to be corrected to an effective distance \cite{Ku05, kumar07}, 
 $L_{ze}$, that can be approach by $L_{ze} \approx L_z -1$.
 The effective distance, due the nanopore flexibility,  
 will oscillate around an average value $\langle L_{ze} \rangle$ 
and the average density will be $\rho = N/(\langle L_{ze} \rangle L^2)$.
Also, it is important to reinforce that, since $N$ is fixed for all simulations,
the distinct values for density are obtained by
the variation in pressure and temperature, and consequently variation in plates separation, $L_z$.

 To analyze the fluid dynamical properties we computed 
 the lateral mean square displacement (MSD) using Einstein relation
 \begin{equation}
 \label{r2}
 \langle [\vec r_{||}(t) - \vec r_{||}(t_0)]^2 \rangle =\langle \Delta \vec r_{||}(t)^2 \rangle=4Dt^\alpha\;,
 \end{equation}
 where $\vec r_{||}(t_0) = (x(t_0)^2 + y(t_0)^2)^{1/2} $ and  $\vec r_{||}(t) = (x(t)^2 + y(t)^2)^{1/2} $
 denote the parallel coordinate of the confined anomalous fluid particle
 at a time $t_0$ and at a later time $t$, respectively.
We should address that the mean square displacement was calculated considering all the particles in the system. 
Nanoconfined fluids assumes a layered structure. Despite this,
the evaluation of $\langle [\vec r_{||}(t) - \vec r_{||}(t_0)]^2 \rangle$ for each layer can lead to a spurious
statistics for the result, since the number of particles in each layer is small and the particles can
move from one layer to another, leading to a poor time average in Eq.~\ref{r2}.
 Depending on the scaling law between $\Delta \vec r_{||}(t)^2$ and $t$ in 
 the limit $t \rightarrow \infty$, different diffusion mechanisms can be identified:
 $\alpha< 1.0$ refers to a subdiffusive regime, with
  $\alpha=0.5$ identifying a single file regime~\cite{Farimani11}. 
 $\alpha=1.0$ stands for a Fickian diffusion whereas $\alpha >1.0$ defines
 the superdiffusive regime, and $\alpha=2.0$ refers to a 
 ballistic diffusion~\cite{Striolo06,Zheng12}.

 In order to define the fluid characteristics at different distances from the nanopore walls, 
 the structure of the fluid layers was analyzed using the radial distribution 
function $g(r_{||})$, defined as
\begin{equation}
\label{gr_lateral}
g(r_{||}) \equiv \frac{1}{\rho ^2V}
\sum_{i\neq j} \delta (r-r_{ij}) \left [ \theta\left( \left|z_i-z_j\right| 
\right) - \theta\left(\left|z_i-z_j\right|-\delta z\right) \right].
\end{equation}

where the Heaviside function $\theta (x)$ restricts the sum of particle pair in a 
slab of thickness $\delta z = 1.0$ close to the wall or $\delta z = 1.0$ away from the walls.

 In all simulations the 
mean variation in the system size induced by the wall fluctuations are smaller than 
$2\%$. Data errors are smaller than the data points and are not shown. The 
data obtained in the equilibration period was not considered for the quantities evaluation.

\section{Results and Discussion}
\label{Results}
         
\begin{figure}[ht]
\begin{center}
\includegraphics[width=8cm]{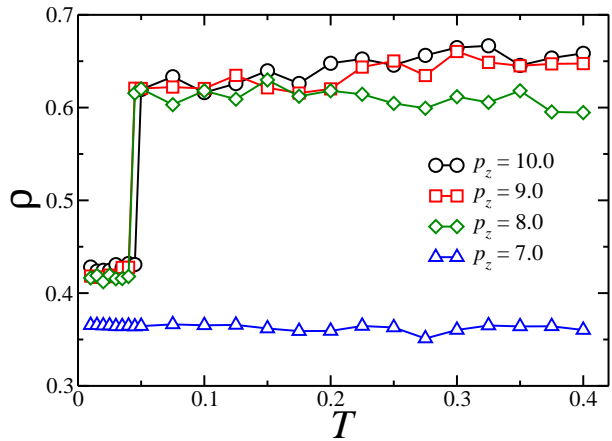}
\end{center}
\caption{$\rho \times T$ phase diagram for the confined anomalous fluid for different isobaric
curves: $p_z = 7.0$, 8.0, 9.0 and 10.0. Errors bars are smaller than the data point.}
\label{dens_mu}
\end{figure}

%
%
%
The thermodynamical behavior of the confined anomalous fluid is 
shown in Fig.~\ref{dens_mu}. The isobaric
curves at lateral pressure $p_z = 7.0$, 8.0, 9.0 and 10.0 show 
distinct behavior. For the smaller pressure,
$p_z = 7.0$, the density as function of the system temperature 
does not vary significatively with the temperature. This result
agrees with our previous findings~\cite{Bordin14a}
that indicates that for flexible walls and pressures $p_z < 6.0$  the 
density varies smoothly with the
temperature. However, for higher values of $p_z$ 
transition from low density to high density is 
observed as the temperature is varied. 
For the pressures values $p_z = $ 8.0 and 9.0 the fluid density
exhibits a jump from the dimensionless density
 $\rho\approx 0.45$ to the density $\rho\approx 0.6$
at $T=0.45$. For $p_z=10.0$ the change in the density occurs 
at the temperature 
$T=0.5$.

This transition is related to a change in the system's conformation. Nanoconfined fluids assume a layered structure~\cite{Nanok09}. 
The number of layers depend  in 
the different nanopores geometries, on the 
nanotube radius and on the plates separation
~\cite{Bordin12b, Bordin13a, Krott13b, Bordin14a, Bordin14b}. Since
the number of particles in our system is fixed, the 
density change observed in the Figure \ref{dens_mu}   implies change in 
the distance between the two plates and consequently
in the number of layers. 

\begin{figure}[ht]
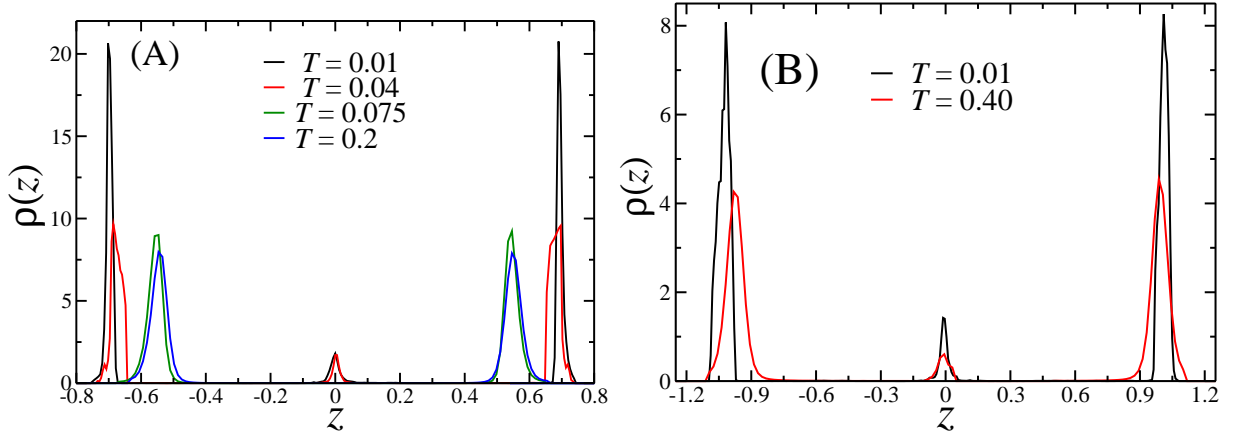

\begin{center}
\includegraphics[width=8cm]{P10hist.eps}
\includegraphics[width=8cm]{P7hist.eps}
\end{center}
\caption{Confined fluid density profile $\rho(z)$ for $p_z=10.0$ and $T = 0.01$, 0.04, 0.075 and 0.2 (A)
and $p_z=7.0$ and $T = 0.01$ and 0.4 (B).}
\label{hist}
\end{figure}

Fig.~\ref{hist}(A) illustrates the 
density distribution versus the distance
between the plates for $p_z=10.0$ at different
temperatures. For low temperatures, the system
forms three layers: two contact layers
and a central layer. For higher temperatures the central
layer melts and the fluid is structured in two contact layers. The behavior for $p_z=8.0$ and 9.0 are similar to the case $p_z=10.0$
and, for simplicity, these results are not shown.

For $p_z=7.0$, the fluid has three layers for
all the temperatures studied as shown in Fig.~\ref{hist}(B).

The transition from low to high density as the temperature
is increased at constant pressure is quite
counterintuitive. Usually 
the increase of density is associated
with decrease of entropy. Here, however, it is
the contrary. This anomalous behavior follows from 
the same mechanism of the increase of 
density at constant pressure, the bulk density anomaly.

\begin{figure}[ht]
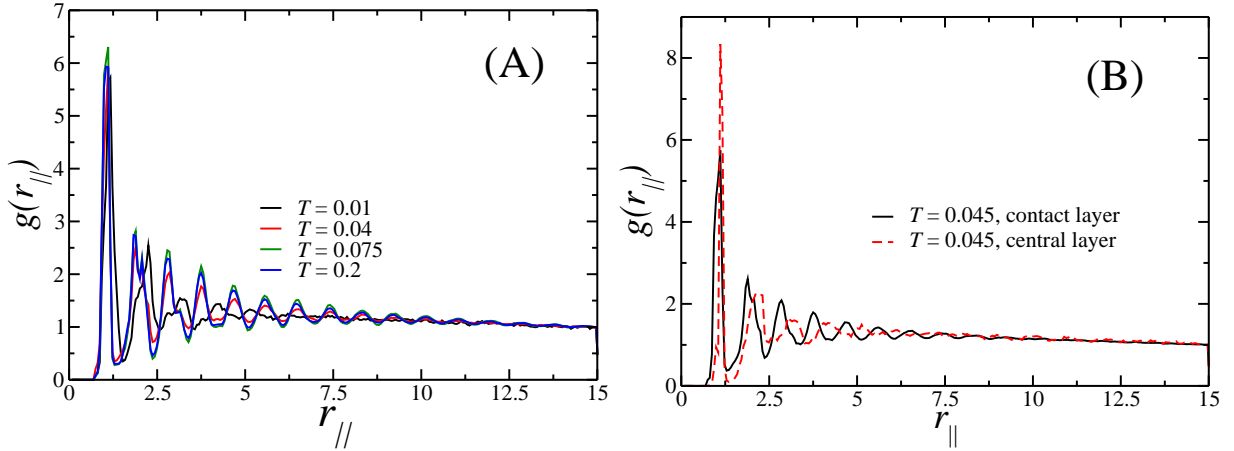

\begin{center}
\includegraphics[width=8cm]{P10gr.eps}
\includegraphics[width=8cm]{P10gr_central.eps}
\end{center}
\caption{(A) Radial distribution function $g(r_{||})$ for the contact layer 
for the confined fluid at $p_z=10.0$ and $T = 0.01$, 0.04, 0.075 and 0.2. (B)
$g(r_{||})$ for the contact and central layer at $p_z=10.0$ and $T = 0.045$.}
\label{gr}
\end{figure}

At low temperatures particles in the same layer
minimize the energy Eq.~(\ref{AlanEq})
by being at a distance
distance  $r_2\approx 2$, the second length in the potential,
as shown in Fig.~\ref{gr}. Because the pressure
is high, the distance between the planes is
smaller than any of the two length scales.

At these low temperatures both the contact planes are quite structured as shown
in Fig.~\ref{gr}(A) while  the central plane is solid-like
as illustrated in Fig.~\ref{gr}(B).
These structures, similarly to the low temperature 
liquid water, have low density but high order and 
consequently low entropy.
As the temperature is increased, the central layer melts. The 
two contact layers approach, being at the 
first length scale, namely  $r_1 \approx 1.2$ 
distance from each other
as shown in Fig.~\ref{hist}. Inside each layer particles
are at $r_2\approx 2$ distant from each other. As the 
temperature is increased, the order inside each layer
decreases  as shown in Fig.~\ref{gr}(A)
and the entropy increases. Therefore, 
the denser system is more entropic similarly with 
what happens with water at the region of
the density anomaly.

\begin{figure}[ht]
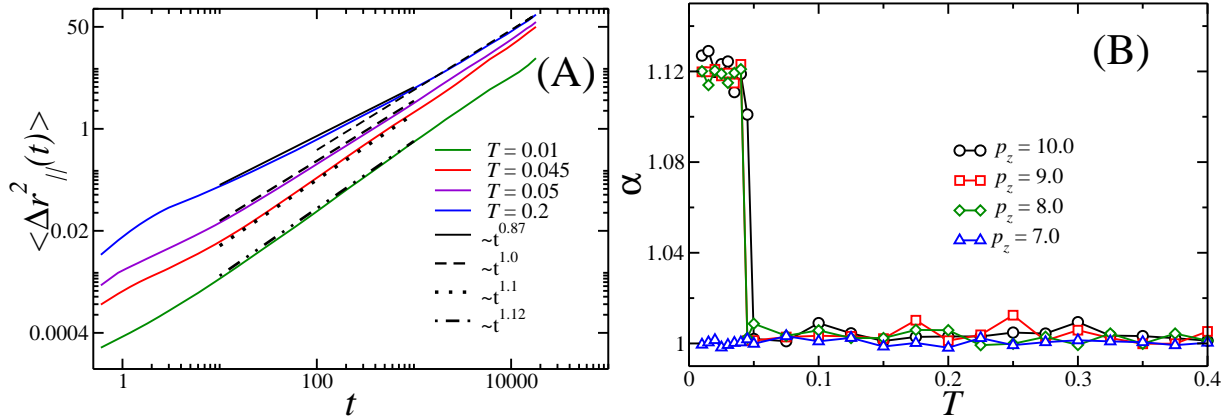

\begin{center}
\includegraphics[width=8cm]{P10inclination.eps}
\includegraphics[width=8cm]{alpha_x_T.eps}
\end{center}
\caption{(A) Lateral mean square displacement $\langle \Delta r^2_{||}(t) \rangle $ as function of simulation
time for external pressure $p_z=10.0$ and $T=0.01$, 0.045, 0.05 and 0.2. Reference curves slopes are
also shown. (B) Temperature dependence of $\alpha$ for $p_z=8.0$, 9.0 and 10.0.}
\label{alpha}
\end{figure}

What is the relation between this layer transition and 
the mobility of the nanoconfined particles? 
The layers structure can provide both a restriction
or an enhancement of the mobility of 
the particles~\cite{Bordin12b, Krott13b, Bordin14b}. 
In order to illustrate this point we
study the
mean square displacement (MSD) as a function 
of time for different temperatures 
and pressures. Fig.~\ref{r2}(A) shows the 
MSD versus time for $p_z=10.0$ and $T=0.01, 0.045, 0.05, 0.2$.
In order to understand the behavior of the mobility in
the framework of Einstein Equation, 
Eq.~(\ref{r2}), the exponent
$\alpha$ is computed 
in all the studied cases. Fickian diffusion if observed for $p_z = 8.0$, 9.0 and 10.0 for  temperatures above the three-to-two layers
transition. This behavior
was obtained after long time simulation. For 
shorter  simulation times the system exhibits an 
apparent  subdiffusive
regime, where $\alpha < 1.0$. This
behavior was also obtained for water
confined in nanotubes~\cite{Zheng12}.
In our case, as in the 
nanotube systems~\cite{Zheng12}, as $t \rightarrow \infty $ 
(see Fig.~\ref{alpha}(A)) the Fickian 
diffusion is recovered.  As example, we show in the purple curve of Fig.~\ref{alpha}(A) the
behavior for $T=0.05$ and $p_z=10.0$.

For low temperatures, below the 
three-to-two layer transition, however, the systems for $p_z=8.0, 9.0$ and $10.0$
exhibit a superdiffusive behavior with $\alpha>1$. 
 
Fig.~\ref{alpha}(B) shows the behavior of $\alpha$ versus
temperature illustrating that the transition temperature from 
non-fickian to fickian regime coincides with the
transition from three-to-two layers with the increase 
of density shown on Fig.~\ref{dens_mu}.

The transition between layers followed by change in the exponent
$\alpha$ was also observed in atomistic models for
water~\cite{Zheng12,Farimani11}. In these
cases, it is not clear is the anomalous diffusion 
are in equilibrium of if they are na artifact of 
the short simulation times In our case,
the coarse grained potential provide us
with an easy way  to perform long simulations and
we can ensure that the system is equilibrated.

\section{Conclusion}
\label{Conclu}

We have studied the behavior of a 
anomalous fluid confined inside a flexible nanopore
at high external pressure. 
Our results show a structural phase transition in the $\rho\times T$ phase diagram
for isobaric curves with $p_z \geq 8.0$. This phase transition corresponds
to a three to two layers transition, and it is associated
to a transition between a superdiffusive regime and a Fickian diffusion.
These results indicates that anomalous fluids, as water, can exhibit
a superdiffusion regime at small temperature and high pressures
associated with the same mechanism that at the bulk generate
the density anomaly

\section{Acknowledgments}

We thanks the Brazilian agencies CNPq, INCT-FCx, and Capes for the financial support.
We also thanks to TSSC - Grupo de Teoria e Simula\c{c}\~{a}o em Sistemas Complexos 
at UFPel for the computer clusters.

\bibliographystyle{aip}

\end{document}